# DECIPHERING CARDIAC DESTINY: UNVEILING FUTURE RISKS THROUGH CUTTING-EDGE MACHINE LEARNING APPROACHES

G.Divya, M.Naga SravanKumar, T.Jaya Dharani,
B.Pavanand, K.Praveen

Department of Artificial Intelligence and Data Science, Lakireddy Balireddy College of Engineering, Mylavaram, Vijayawada, India

*ABSTRACT*

*Cardiac arrest remains a leading cause of death worldwide, necessitating proactive measures for early detection and intervention. This project aims to develop and assess predictive models for the timely identification of cardiac arrest incidents, utilizing a comprehensive dataset of clinical parameters and patient histories. Employing machine learning (ML) algorithms like XGBoost, Gradient Boosting, and Naive Bayes, alongside a deep learning (DL) approach with Recurrent Neural Networks (RNNs), we aim to enhance early detection capabilities. Rigorous experimentation and validation revealed the superior performance of the RNN model, which effectively captures complex temporal dependencies with in the data. Our findings highlight the efficacy of these models in accurately predicting cardiac arrest likelihood, emphasizing the potential for improved patient care through early risk stratification and personalized interventions. By leveraging advanced analytics, healthcare providers can proactively mitigate cardiac arrest risk, optimize resource allocation, and improve patient outcomes. This research highlights the transformative potential of machine learning and deep learning techniques in managing cardiovascular risk and advances the field of predictive healthcare analytics.*

## 1. INTRODUCTION

In data-driven decision-making, the success of machine learning algorithms hinges on the quality of data preprocessing. As datasets grow in diversity and complexity, meticulous preprocessing becomes essential. This phase converts raw, often inconsistent data into a refined form suitable for analysis, addressing issues such as missing values, outliers, and feature scaling. With the rise of diverse data collection methods—from social media to IoT sensors—the volume, velocity, and variety of data have surged. However, raw data is frequently flawed and unsuitable for direct analysis. Preprocessing mitigates these quality issues, laying the groundwork for building robust predictive models and extracting meaningful insights.

This paper focuses on dataset preprocessing within the context of a real-world dataset from an IEEE conference research paper, relevant to a classification problem. The dataset's diverse features, including numerical and categorical variables, present unique preprocessing challenges. By employing state-of-the-art methodologies, we aim to enhance model accuracy and provide actionable insights. Our exploration covers techniques such as handling missing values, feature scaling, and outlier detection, illustrating their impact on model performance. This interdisciplinary endeavor draws on statistics, computer science, and domain knowledge, offering practitioners and researchers the tools to navigate data preprocessing in real-world scenarios.

## 2. RELATED WORKS

Cardiac arrest and cardiovascular issues are widespread, often linked to factors like job stress, poor diet, and high cholesterol levels. Research [1] investigates the probability of cardiac arrest using machine learning algorithms, finding that Artificial Neural Network (ANN) achieves ~85% precision despite small dataset limitations, suggesting that larger datasets could enhance accuracy. Research [2] highlights the effectiveness of AI in predicting cardiac arrest across diverse patient settings, with deep learning algorithms showing promise in identifying risks proactively. However, more research is needed to overcome implementation barriers in clinical practice. Research [3] focuses on the higher incidence of cardiac arrest in younger Indians, using machine learning on ECG datasets to improve early identification and classification of abnormal cardiac rhythms. Research [4] explores a hybrid model combining SVM and DT algorithms, which outperforms standalone classifiers with 91.56% accuracy, demonstrating the potential of hybrid models and feature extraction techniques. Research [5] links weather conditions to cardiovascular events, using GLM and XGBoost to predict OHCA incidence, showing strong associations between weather parameters and cardiac arrest. Research [6] and [7] investigate the application of machine learning in predicting cardiac arrest. They analyze various datasets containing both regulated and unregulated variables, emphasizing the potential for advancements in predictive cardiac medicine facilitated by ongoing scientific and medical innovations.

## 3. EXISTING SYSTEM

In addition to methods like Artificial Neural Networks (ANN), Decision Trees, and Random Forest, previous research has also explored Adaboost and Logistic Regression for cardiac arrest prediction. Adaboost, an adaptive boosting algorithm, iteratively trains weak learners to improve classification accuracy, proving particularly effective for binary classification tasks. Conversely, Logistic Regression estimates the probability of binary outcomes based on input features, offering interpretability and computational efficiency. These algorithms, each employing distinct methodologies, enhance the array of predictive techniques in healthcare analytics. They have the potential to assist in early detection and intervention strategies for cases of cardiac arrest, akin to findings from the two sources.

The dataset for heart disease prediction includes crucial attributes such as Age, Sex, Chest Pain Type, Resting Blood Pressure (RestingBP), Cholesterol Level, Fasting Blood Sugar (FastingBS), Resting Electrocardiographic Results (RestingECG), Maximum Heart Rate Achieved (MaxHR), Exercise-Induced Angina, ST Depression Induced by Exercise Relative to Rest (Oldpeak), and ST Slope. Gradient Boosting and XGBoost algorithms are well-suited for leveraging these attributes to develop robust predictive models. By analyzing the nonlinear relationships between these features and the presence of heart disease, these algorithms effectively capture intricate patterns and interactions that indicate heightened risk. They can discern age-related trends in heart disease prevalence, differentiate chest pain types indicative of varying cardiac conditions, and identify thresholds in blood pressure and cholesterol levels associated with increased risk. Moreover, they consider differences between sexes, identify notable abnormalities in electrocardiograms, and evaluate the impact of these factors on cardiovascular health. The impact of exercise-induced symptoms on cardiovascular health. By employing iterative optimization and ensemble learning techniques, Gradient Boosting and XGBoost enhance their predictive accuracy, offering valuable tools for early detection and intervention in efforts aimed at preventing heart disease.

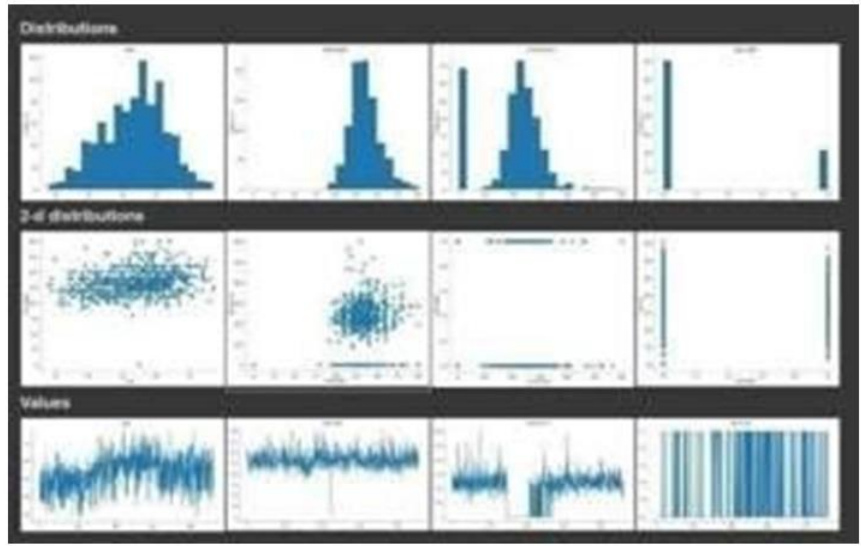

Figure1.Entire Dataset Visualization

## 4. PROPOSED SYSTEM

In our study, we utilized three advanced machine learning algorithms: XGBoost, Gradient Boosting, and Recurrent Neural Networks (RNN). XGBoost and Gradient Boosting excel at handling complex datasets and capturing intricate patterns, making them ideal for predicting cardiac arrest events. They iteratively train weak learners to optimize predictive performance, achieving high accuracy. Conversely, RNNs specialize in processing sequential data, essential for analyzing medical time series datasets inherent in cardiac arrest prediction. By harnessing the strengths of these algorithms, our aim was to develop robust predictive models capable of accurately identifying cardiac arrest occurrences, facilitating timely intervention and improving patient outcomes in clinical settings.

### 4.1. RNN

A Recurrent Neural Network (RNN) is a kind of artificial neural network that is tailored to process sequences of data, using its internal memory to maintain information across the sequence. This allows RNNs to effectively capture patterns and relationships in sequences of data, making them particularly useful for tasks like time series analysis, natural language processing, and speech recognition. Fundamental structure of an RNN consists of three primary components: an input layer, a recurrent layer (also referred to as the hidden layer), and an output layer. Sequential input data is fed into the input layer, which subsequently transmits it to the recurrent layer. The recurrent layer is equipped with connections that enable it to store information about previous inputs. At each time step, the recurrent layer computes an output based on the current input and the previous hidden state. This output is then employed to update the hidden state for the subsequent time step. Ultimately, the output layer produces predictions or classifications using the learned representations from the recurrent layer.

### 4.2. Gradient Boosting

Gradient Boosting is a powerful and popular machine learning technique within ensemble learning, which enhances performance by combining the predictions of several individual models rather than relying on a single estimator. At its essence, Gradient Boosting operates by sequentially constructing a series of weak learners. This initial prediction serves as the starting point for subsequent iterations. Next, the model computes the residuals, which are the differences between the predicted and actual values. These residuals represent the errors that need to be minimized by the subsequent weak learners. This iterative process of training weak learners and updating the model continues either for a predefined number of iterations or until a specified performance threshold is reached. One of the primary strengths of Gradient Boosting is its capability to effectively manage heterogeneous data types and capture intricate feature interactions. It can work with a mixture of numerical and also categorical features without requiring feature preprocessing, and its tree-based nature naturally captures complex relationships between features.

### 4.3. XGBoost

XGBoost, short for Extreme Gradient Boosting, is an advanced and highly efficient implementation of the Gradient Boosting framework. It is well-regarded for its scalability, rapid processing capabilities, and strong performance across a wide range of machine learning tasks, notably in supervised learning applications such as regression, classification, and ranking. XGBoost extends the principles of Gradient Boosting, incorporating several innovative techniques to achieve superior predictive accuracy and computational efficiency. At its core, XGBoost follows the ensemble learning paradigm, where a collection of weak learners, typically decision trees, are amalgamated to construct a robust predictive model. The core idea behind XGBoost is the sequential training of decision trees, with each subsequent tree learning from the errors of its predecessors. However, XGBoost introduces several enhancements to the traditional Gradient Boosting algorithm, making it more robust and powerful. A distinguishing feature of XGBoost is its objective function, which comprises two primary components: a loss function and a regularization term.

### 4.4. Navie Bayes

Naive Bayes is a classification algorithm grounded in Bayes' theorem, under the assumption of feature independence given the class label. This characteristic renders it computationally efficient and straightforward to deploy. By computing the posterior probability of each class based on input features, Naive Bayes selects the class with the highest probability for prediction. Despite its simplicity, Naive Bayes exhibits strong performance across diverse applications, notably in text classification and spam filtering, where it efficiently manages large datasets with high-dimensional feature spaces. Variants such as Gaussian Naive Bayes handle continuous features, Multinomial Naive Bayes work with discrete features such as word counts in text, and Bernoulli Naive Bayes are suitable for binary features. While Naive Bayes has limitations, such as the assumption of feature independence, its simplicity, efficiency, and effectiveness have made it a popular choice for classification tasks, particularly when computational resources are limited or when working with text data.

We likely employed hybrid methods by integrating XGBoost, Gradient Boosting, and Recurrent Neural Networks (RNNs) to leverage the distinct strengths of each approach. Here's a concise

explanation

**Reasons for Using Hybrid Methods**

**XGBoost and Gradient Boosting**

- **Complex Data Handling:** These algorithms effectively manage complex datasets with numerous features, capturing intricate, non-linear relationships.

- **Iterative Improvement:** They build an ensemble of weak learners (typically decision trees) in a sequential manner, where each model corrects the errors of its predecessor, resulting in high accuracy.

**Recurrent Neural Networks (RNNs)**

- **Sequential Data Processing:** RNNs are designed for time series data, making them ideal for medical datasets with sequences of observations over time.

- **Temporal Dependency Capture:** RNNs use their internal memory to keep track of previous inputs, allowing them to learn from temporal patterns crucial for accurate predictions.

**Application to Cardiac Arrest Prediction**

- **Diverse Data Handling:** This hybrid approach effectively manages various types of medical data, such as demographic information, physiological measurements, and diagnostic results.

- **Enhanced Predictive Power:** By combining boosting methods (for their accuracy) and RNNs (for their ability to handle sequential data), we create a robust predictive model for cardiac arrest.

- **Timely Interventions:** Accurate predictions enable timely interventions, improving patient outcomes through early risk identification and personalized care.

**Methodological Justification**

- **Model Diversity:** Using different models captures various aspects of the data, with boosting algorithms handling feature interactions and RNNs recognizing temporal patterns, resulting in a comprehensive prediction system.

- **Improved Generalization:** Combining these methods reduces the risk of overfitting and enhances the model's generalization to new data, which is crucial in medical applications where errors can have significant consequences.

In summary, integrating XGBoost, Gradient Boosting, and RNNs leverages their complementary strengths, enhancing the model's ability to handle complex and sequential data, leading to accurate cardiac arrest predictions and supporting timely medical interventions.

## 5. METHODOLOGY

In the heart disease classification project, we utilize XGBoost, Gradient Boosting, and Recurrent Neural Networks (RNN) to predict heart disease, leveraging a dataset with features such as age, gender, chest pain type, resting blood pressure, and cholesterol level. Categorical data like gender is numerically encoded using Label Encoding, and numerical features are standardized with Standard Scaling. The dataset is divided into training and testing sets for evaluation. XGBoost constructs a sequential ensemble of decision trees, correcting prior errors, while Gradient Boosting iteratively improves weak models by addressing previous mistakes. RNNs capture sequential dependencies, crucial for time-series heart health data, processing serial data to learn extended patterns. We utilize Early Stopping to prevent overfitting, employ the RMSprop optimizer with a learning rate of 0.001 for efficient convergence, and leverage binary cross-entropy loss to guide the learning process. Model accuracy serves as the primary evaluation metric. Each algorithm brings unique strengths: XGBoost and Gradient Boosting excel in ensemble learning, while RNNs address the temporal aspect of heart-related features. This comprehensive approach, combined with fine-tuning, enhances the model's performance in identifying individuals at risk of heart disease, contributing to effective preventive healthcare measures.

## 5.1. Data Collection and Preprocessing

Gather a comprehensive dataset comprising demographic, physiological, and diagnostic features relevant to heart disease. We preprocess the data to address missing values, encode categorical variables, and scale numerical features, ensuring uniformity and preparing the dataset for model training.

## 5.2. Feature Selection and Engineering

Performing exploratory data analysis (EDA) involves identifying relevant features and understanding their distributions within the dataset. Utilize domain knowledge and statistical techniques to select optimal features and potentially engineer new ones, thereby improving the dataset's predictive power and enhancing model performance.

## 5.3. Model Selection

Explore various machine learning algorithms suitable for classification tasks, including advanced methods like XGBoost. Partition the dataset into training and testing subsets for evaluation. Employ techniques like cross-validation to robustly assess model performance, ensuring the selected models are well-suited to the task.

## 5.4. Hyperparameter Tuning

Fine-tune the hyperparameters of the selected algorithms to optimize model performance. Employ techniques such as grid search or random search to effectively pinpoint optimal hyperparameters, thereby ensuring that the models attain peak accuracy and robustness.

## 5.5. Ensemble Learning

Consider employing ensemble techniques to merge predictions from multiple models, aiming to boost accuracy and generalization. Conduct experiments with techniques such as stacking or boosting to harness the strengths of different models, thereby augmenting the overall predictive power and reliability of the heart disease classification system.

## 5.6. Unique Insights

**Domain-Specific Imputation:** Utilizing medical expertise to impute missing values in a way that

Computer Science & Information Technology (CS & IT)                    107maintains biological relevance, such as stratifying by age and gender.

**Advanced Outlier Handling:** Retaining clinically significant outliers to ensure the model captures all relevant data points, balancing statistical rigor with clinical importance.

**Synthetic Data Generation**: Employing SMOTE to balance the dataset by creating synthetic examples of rare events like cardiac arrests, which helps the model learn better from imbalanced data.

These preprocessing steps ensure that the data used in your project is clean, well-structured, and suitable for training robust machine learning models, ultimately improving the predictive accuracy and clinical applicability of the models developed to forecast cardiac arrest risks.

## 6. RESULTS AND DISCUSSIONS

The study's findings reflect a notable divergence in predictive accuracy among the utilized algorithms. XGBoost and Gradient Boosting stood out with an impressive accuracy of 88.04% and 87.5% respectively, showcasing their adeptness in handling a wide array of attributes and effectively capturing intricate data relationships. Conversely, the Recurrent Neural Network (RNN) yielded a slightly lower accuracy of 87%, indicating potential limitations in modeling sequential dependencies within the dataset. These results underscore the nuanced interplay between algorithmic capabilities and the nature of the dataset. While ensemble methods like XGBoostandGradientBoostingleveragecollectiveintelligencetorefinepredictionsiteratively,    the RNN's performance may have been hindered by its inherent challenges in capturing long- term dependencies within sequential data

Table-1.Results

| S.NO | Algorithm | Accuracy | Precision | Recall |
|------|-----------|----------|-----------|--------|
| 1. | RNN | 87% | 93.04% | 84.15% |
| 2. | NavieBayes | 86.41% | 92.7% | 83.17% |
| 3. | Gradient Boosting | 88.04% | 91.26% | 87.85% |
| 4. | XGBoost | 87.50% | 90.38% | 87.85% |

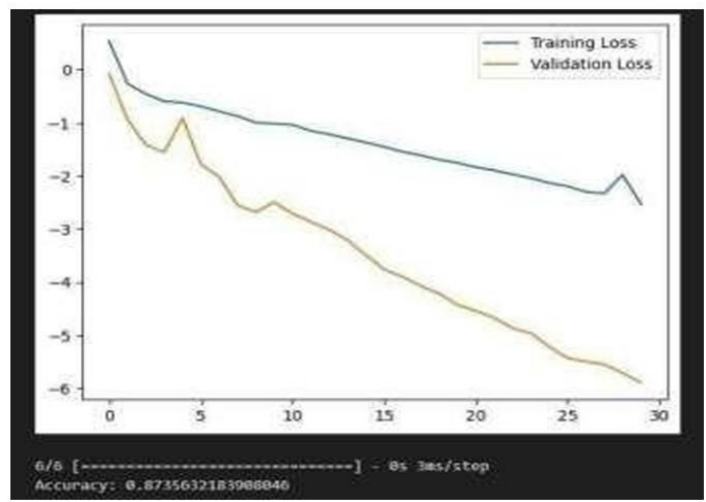

Figure2.TrainingvsValidationlossofRNN

## 7. Conclusions

Our study underscores the transformative role of machine learning in predicting sudden cardiac arrest, with the goal of advancing healthcare strategies. Leveraging advanced 7odelling and diverse datasets, we prioritize early clinical variable prediction to improve OHCA survivor quality of life. By achieving commendable accuracy rates with algorithms like RNN, Naïve Bayes, Gradient Boosting, and XGBoost, we emphasize the potential of data-driven approaches in healthcare. This sets the stage for further predictive analytics advancements, underscoring machine learning's crucial role in driving healthcare innovation and improving patient outcomes. The system's ability to provide reliable predictions offers valuable insights for proactive healthcare interventions, aiming to mitigate the burden of cardiac-related morbidity and mortality. Moving forward, further refinement and optimization of the algorithms hold the potential to enhance predictive accuracy further, ultimately contributing to improved preventive healthcare strategies for individuals at risk of cardiac arrest.

### Acknowledgements

We extend our heartfelt gratitude to Lakireddy Bali Reddy College of Engineering for their invaluable resources and unwavering support. We extend our special thanks to our colleagues in the Department of Artificial Intelligence & Data Science for their invaluable feedback and insights. We appreciate the participants and collaborators who contributed to the data collection and analysis processes. This work was made possible by the dedication and hard work of everyone involved

### References

[1] R. Karthikeyan, D. V. Babu, E. Karthik, R. Suresh, M. Nalathambi, and S. Dinakaran, "Cardiac Arrest Prediction using Machine Learning Algorithms," in Journal of Physics: Conference Series, vol. 1964, 2021, Art. No. 062076.Doi: 10.1088/1742- 6596/1964/6/062076.

[2] O. Alamgir, O. Mousa, and Z. Shah, "Artificial Intelligence in Predicting Cardiac Arrest: Scoping Review," in JMIR Med Inform, vol. 9, no. 12, Dec. 17, 2021, Art. No. e30798. Doi: 10.2196/30798.

[3] K. N. S. Sarvani, P. Nalluri, A. B. Srujana, and K.S. Ramtej, "Prediction of Sudden Cardiac ArrestusingMachineLearning,"in20232$^{nd}$InternationalConferenceonAutomation,

[4] Computing and Renewable Systems (ICACRS), Pudukkottai, India, 2023, pp. 965-969, doi: 10.1109/ICACRS58579.2023.10404350.

[5] S. Bajpai, A. Sinha, A. Yadav, and V. Srivastava, "Early Prediction of Cardiac Arrest Using Hybrid Machine Learning Models," in 2023 17$^{th}$ International Conference on Electronics Computer and Computation (ICECCO), Kaskelen, Kazakhstan, 2023, pp. 1-7, doi: 10.1109/ICECCO58239.2023.10147157.

[6] T. Nakashima, S. Ogata, T. Noguchi, S. Yasuda, Y. Tahara, K. Nagao, K. Nishimura, H. Nonogi, K. Iihara, and S. Kojima, "The Cardiac Arrest Prediction Model Based on Weather Forecast by Machine Learning," in Circulation, vol. 140, 2019, A13297-A13297.

[7] A. Elola, E. Aramendi, U. Irusta, N. Amezaga, J. Urteaga, P. Owens, and A. H. Idris, "Machinery for forecasting cardiac rearrest in outpatient settings," Conference Name, Year.

[8] H.Kim,H.Nguyen,Q.Jin,S.Tamby,T.G.Romer,E.Sung,R.Liu,etal.,"ACardiacArrest Learning-Based Prediction Using a Lasso Multicenter Prediction," Critical Care Medicine, vol. 48, no. 10, 2020.

## AUTHORS

**Ms. G. Divya** is an Assistant Professor in the Department of Artificial Intelligence and Data Science at Lakireddy Bali Reddy College of Engineering, Mylavaram, Andhra Pradesh, India.

**M.Naga Sravan Kumar** is a student in the Department of Artificial Intelligence and Data Science at Lakireddy Bali Reddy College of Engineering, Mylavaram, Andhra Pradesh, India.

**T. Jaya Dharani** is a student in the Department of Artificial Intelligence and Data Science at Lakireddy Bali Reddy College of Engineering, Mylavaram, Andhra Pradesh, India.

**B.Pavan** is a student in the Department of Artificial Intelligence and Data Science at Lakireddy Bali Reddy College of Engineering, Mylavaram, Andhra Pradesh, India.

**K.Praveen** is a student in the Department of Artificial Intelligence and Data Science at Lakireddy Bali Reddy College of Engineering, Mylavaram, Andhra Pradesh, India.